\documentstyle[12pt,epsfig]{article}

\newlength{\dinwidth}
\newlength{\dinmargin}
\def\lapproxeq{\lower .7ex\hbox{$\;\stackrel{\textstyle
<}{\sim}\;$}}
\def\gapproxeq{\lower .7ex\hbox{$\;\stackrel{\textstyle
>}{\sim}\;$}}

\def\beq{\begin{equation}}
\def\eeq{\end{equation}}
\def\bea{\begin{eqnarray}}
\def\eea{\end{eqnarray}}

\def\MeV{\rm MeV}
\def\GeV{\rm GeV}
\def\msb{\overline{\rm MS}}

\begin{document}
\titlepage
\begin{flushright}
IPPP/06/02 \\
DCPT/06/04 \\
Cavendish-HEP-2006/02 \\

\end{flushright}

\vspace*{0.5cm}

\begin{center}
{\Large \bf The role of $F_L(x,Q^2)$ in parton analyses. }

\vspace*{1cm}
\textsc{A.D. Martin$^a$, W.J. Stirling$^a$
and R.S. Thorne$^{c,}$\footnote{Royal Society University Research Fellow.}} \\

\vspace*{0.5cm} $^a$ Institute for Particle Physics Phenomenology,
University of Durham, DH1 3LE, UK \\
$^c$ Cavendish Laboratory, University of Cambridge, \\ JJ Thomson Avenue,
Cambridge, CB3 0HE, UK
\end{center}

\vspace*{0.5cm}

\begin{abstract}
We investigate the effect of the structure function $F_L$ in global parton analyses of deep inelastic 
and related hard scattering data.  We perform NLO and NNLO analyses which include the reduced cross
section HERA data at high $y$, as well as earlier direct measurements of $F_L$.  We find that the NNLO analysis
gives a better description of $F_L$ at low $x$ than that performed at NLO.  Nevertheless the data show evidence
of the need of further contributions to $F_L$, which may be of higher-twist origin.  
We study such corrections both phenomenologically and theoretically via a
renormalon approach.  The higher-twist corrections extracted from a successful 
fit to the data are in general agreement with
the theoretical expectations, but there is still room for alternative 
theoretical contributions, particularly at
low $x$ and $Q^2$.  The importance of future measurements of $F_L$ is emphasized.
\end{abstract}

\vspace*{0.5cm}
\section{Introduction}

The cross-section for deep-inelastic charged lepton$-$proton scattering depends on the two independent 
structure functions $F_2(x,Q^2)$ and $F_L(x,Q^2)$. The former is dominated by 
the quark parton distributions, and the latter, in principle, by the gluon
distribution (except at high $x$). However, $F_2(x,Q^2)$ is found to be much larger 
than $F_L(x,Q^2)$.  Moreover, they appear in the cross-section in the 
combination  
\beq
\label{eq:sigred}
\tilde \sigma(x,Q^2) = F_2(x,Q^2) -\frac{y^2}{1+(1-y)^2}F_L(x,Q^2),
\eeq
where $y=Q^2/xs$. The quantity $\tilde \sigma(x,Q^2)$ has become known as the 
``reduced cross-section''. Since $y\ll 1$ in most of the kinematic range,  
$\tilde \sigma(x,Q^2)$ is effectively the same as $F_2(x,Q^2)$. However, at HERA, 
for the lowest $x$-values at given $Q^2$, the value of $y$ can become as large as 
$0.7-0.8$, and the effect of $F_L(x,Q^2)$ becomes apparent \cite{H1FL}. 
This is seen in 
the data as a flattening of the growth of $\tilde \sigma(x,Q^2)$ as $x$ decreases to very small values
(for fixed $Q^2$), leading eventually to a turnover. Hence, when analysing the HERA
structure function data it is particularly important to fit any theoretical 
prediction to the measured $\tilde \sigma(x,Q^2)$, rather than to  
model-dependent extracted values of $F_2(x,Q^2)$ \cite{ThorneFL}. Indeed, important lessons may be learned by placing 
particular emphasis on the data at very high $y$. In this paper we
examine the impact of the contribution from $F_L(x,Q^2)$ in this region.

As well as this very low $x$ HERA data, we will also study the impact of the much 
higher $x$ direct measurements of $F_L(x,Q^2)$, which were made by SLAC \cite{SLAC}, 
BCDMS \cite{BCDMS} and NMC \cite{NMC}
by measuring the cross-section at a variety of values of $y$. In this region 
the contribution from both quarks and gluons is obviously important, as 
indeed it turns out to be at the low-$x$ values. It has recently been proposed 
to make a direct determination of $F_L$ at the low-$x$ values accessible at HERA by 
making some measurements of the cross-sections with lowered beam energy
\cite{Klein}. There is also a possibility of a measurement associated with 
eRHIC \cite{Caldwell}. 
We conclude by discussing the importance of such future measurements.

\section{Perturbative Stability of $F_L(x,Q^2)$}

\begin{figure}
\begin{center}
\centerline{\epsfxsize=0.7\textwidth\epsfbox{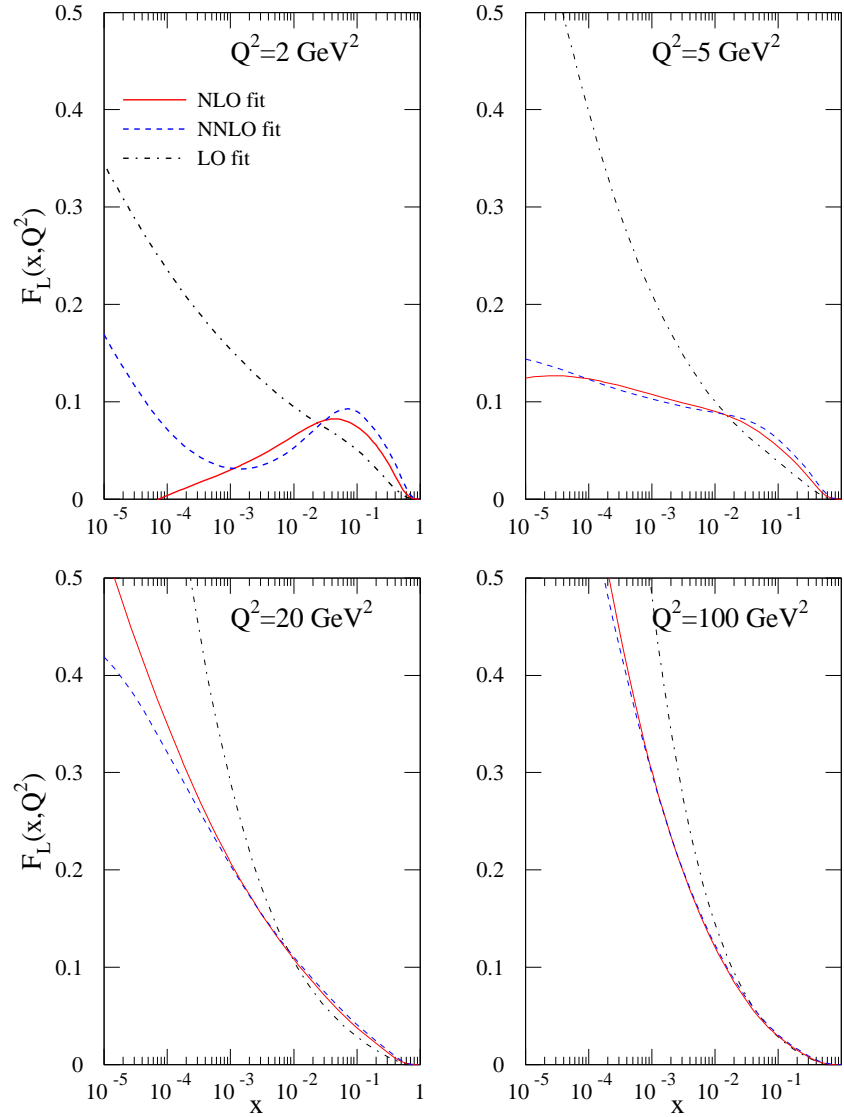}}
\vspace{-0.3cm}
\caption{$F_L(x,Q^2)$ predicted from the global fit at LO, 
NLO and NNLO.}
\vspace{-1cm}
\label{fig:stab}
\end{center}
\end{figure}

We have recently been able to make accurate and reliable predictions of 
$F_L(x,Q^2)$ up to NNLO in perturbative QCD \cite{NNLOs,NNLOfl}. 
The procedure is to first determine the parton distribution functions from a
global fit to the available deep inelastic and related hard scattering data,
without the inclusion of any $F_L$ data.  For instance, the gluon distribution is
constrained at small $x$ by measurements of $\partial F_2/\partial {\rm ln}Q^2$, and at
large $x$ by Tevatron jet data.  The extracted partons are then used to predict $F_L$.
In this way we can study the 
perturbative stability of this fundamental quantity as one increases the 
order of the calculation. 
The results obtained are shown in 
Fig.~\ref{fig:stab}.\footnote{The NNLO variable-flavour number scheme 
\cite{VFNS} has been used in the NNLO global analysis. The details of the 
results are sensitive to this updated treatment of heavy flavours.}
It is immediately clear that at NLO there is a serious problem with 
$F_L(x,Q^2)$ at the lowest values of $x$ and $Q^2$ with it becoming 
(unphysically) negative. This is a reflection of the behaviour of the gluon 
distribution at the same order. However, the NNLO coefficient functions for 
$F_L(x,Q^2)$, $C_{Lg,q}^{\rm NNLO}$, turn out to be large and positive at small $x$ (for both 
quarks and gluons). In detail
\beq
F_L=\alpha_S(C_{Lg}^{\rm LO}+\alpha _S C_{Lg}^{\rm NLO}+\alpha _S^2 C_{Lg}^{\rm NNLO}+...) \otimes g~~+~~g \to q
\eeq
where up to NLO the shape of $F_L(x,Q^2)$ is dominated by that of the 
partons, particularly the gluon at low $x$. ($C_{Lg}^{\rm NLO}$ is divergent at
small $x$, but the $1/x$ term has a very small, negative coefficient.)
The NNLO longitudinal coefficient function $C^{NNLO}_{Lg}(x)$ is given by
\begin{equation}
C^{\rm NNLO}_{Lg}(x) = n_f \biggl(\frac{1}{4\pi}\biggr)^3
\biggl(\frac{409.5\ln(1/x)}{x} -\frac{2044.7}{x}-\cdots\biggr).
\end{equation} 
There is clearly a significant positive contribution at very small $x$, 
$x\ll 0.01$, and 
this counters the decrease in the small-$x$ gluon. 
Hence, even though the gluon is even more negative at
small $x$ and $Q^2$ at NNLO than it is at NLO, the prediction for $F_L(x,Q^2)$
has become positive. Indeed, the effect of the NNLO coefficient functions is so 
important, at low $x$ and $Q^2$, that $F_L(x,Q^2)$ starts to increase as $x$ 
decreases below about $10^{-3}$. At higher $Q^2$, i.e. $Q^2 \lapproxeq 5~\GeV^2$, the NLO and 
NNLO predictions are not too dissimilar at small $x$, though the current very 
close agreement at $Q^2=5~\GeV^2$ is coincidental.     

\begin{figure}
\begin{center}
\centerline{\epsfxsize=0.6\textwidth\epsfbox{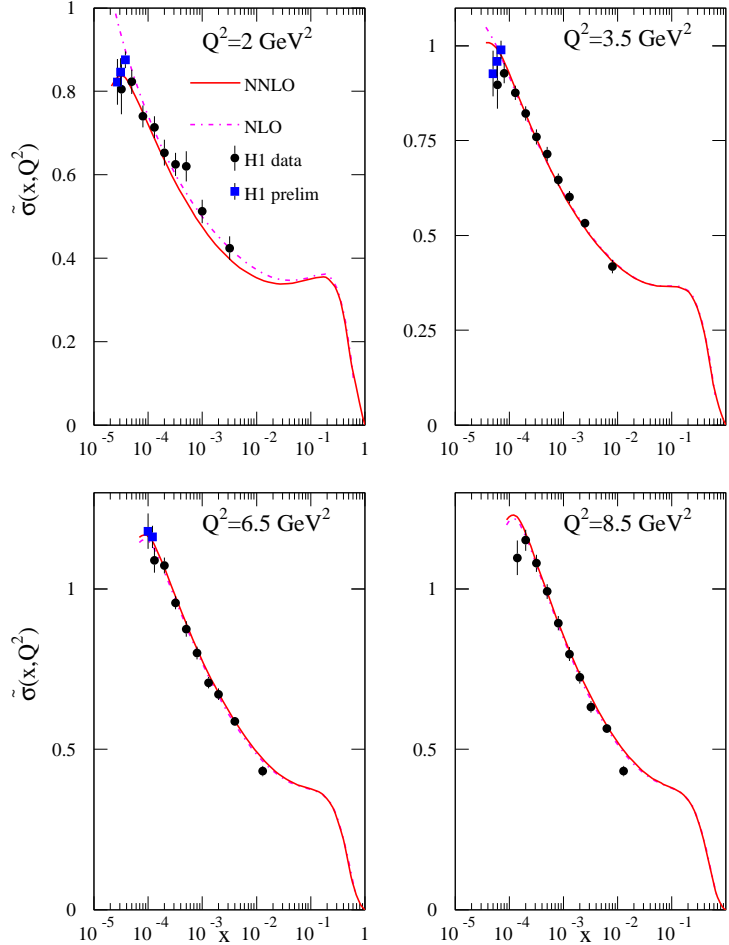}}
\vspace{-0.3cm}
\caption{The consistency check of  $F_L(x,Q^2)$ for the 
NLO and NNLO MRST fits.}
\vspace{-1cm}
\label{fig:sigred}
\end{center}
\end{figure}

This predicted increase in $F_L(x,Q^2)$ is very important for the 
comparison with the high-$y$ HERA $\tilde \sigma$ data. The very small, or even negative, values 
at NLO mean that there is no turn-over in the theoretical curves to 
accompany that in the data, as seen in Fig.~\ref{fig:sigred}. However, the discrepancy
is cured at NNLO, and the comparison with $\tilde \sigma(x,Q^2)$ 
is quite good. In fact, if the NNLO $\msb$ gluon distribution were positive 
definite at input ($Q_0^2=1~\GeV^2$) the resulting $F_L(x,Q^2)$ would be 
rather too large at the smallest $x$ and $Q^2$ and the turnover in
$\tilde \sigma(x,Q^2)$ would be too large. (Also the shape of $F_L(x,Q^2)$ 
with $Q^2$ at small $x$ would be very strange, growing quickly as $Q^2$
decreases.) This illustrates the fact that the small-$x$ gluon is a scheme-dependent,
unphysical quantity, which in the $\msb$ scheme is very unlike the physical 
$F_L(x,Q^2)$.

Bearing this in mind, we attempted to construct a more 
``physical'' definition for the small-$x$ gluon, in a similar spirit to that we used recently 
for the high-$x$ gluon \cite{MRST04}. 
Explicitly, we invented a 
scheme where the gluon was defined by
\beq
\label{eq:phys}
\tilde g^{F_L}(x,Q^2) = \biggl(\delta(1-x)+\frac{\alpha_S \tilde C^{\rm NLO}_{Lg}
+\alpha^2_S \tilde C_{Lg}^{\rm NNLO}}{C^{\rm LO}_{Lg}}\biggr)\otimes g^{\msb}(x,Q^2),
\eeq  
where $\tilde C_{Lg}^{\rm NLO}$ and $\tilde C^{\rm NNLO}_{Lg}$ are functions identical
to the NLO and NNLO coefficient functions in the small-$x$ limit, but modified
at high-$x$ so that momentum is conserved in the transformation between 
schemes. If the exact coefficient functions were used in Eq.(\ref{eq:phys})
then the ``physical'' definition of the gluon, $\tilde g^{F_L}(x,Q^2)$ 
would be guaranteed to be the same shape as $F_L(x,Q^2)$ at small $x$, and 
hence would have a genuine physical interpretation in this scheme. However,
performing fits where we defined as input $\tilde g^{F_L}(x,Q_0^2)$ and 
transformed to the $\msb$ scheme (as we defined the high-$x$ gluon in DIS 
scheme before transforming to the $\msb$ scheme in \cite{MRST04}) we found that
$\tilde g^{F_L}(x,Q^2)$ still tended to be valence-like or even negative at 
small $x$. This is because the exact $C^{\rm NLO}_{Lg}$, and particularly
$C^{\rm NNLO}_{Lg}$, do not have zero first moment, i.e. are not momentum 
conserving, and in fact using the real coefficients in  Eq.(\ref{eq:phys})
would leads to a considerably larger value of $\tilde g^{F_L}(x,Q^2)$ than momentum
conserving functions can. Hence, we conclude that it is not easy to devise
a simple scheme where the low-$x$ gluon behaves like $F_L(x,Q^2)$, but where 
the interpretation in terms of parton distribution functions as probabilities 
is clearly maintained.  

\begin{figure}
\begin{center}
\centerline{\epsfxsize=0.7\textwidth\epsfbox{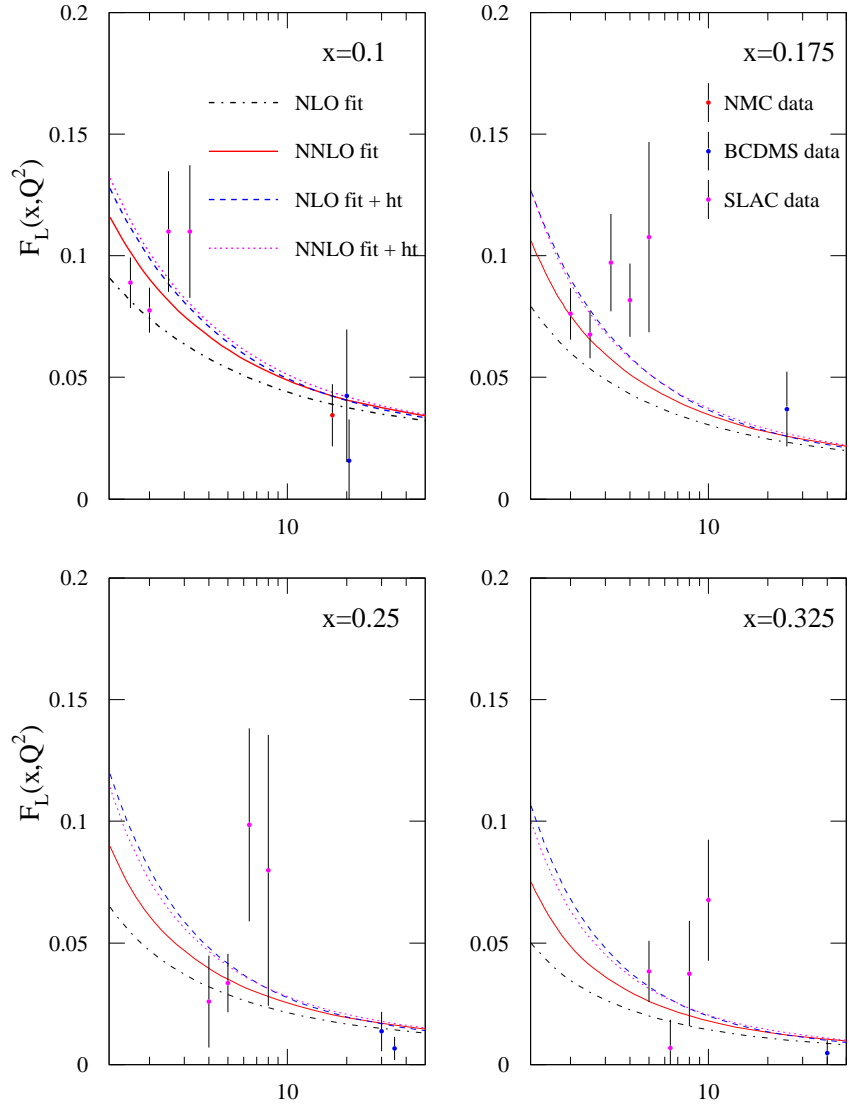}}
\vspace{-0.3cm}
\caption{The comparison with the direct data on $F_L(x,Q^2)$ at NLO and NNLO; 
and also after including higher twist contributions as described in the text.}
\vspace{-1cm}
\label{fig:direct}
\end{center}
\end{figure}

We conclude that even though the NNLO prediction for $F_L(x,Q^2)$ is much 
better than that at NLO, there are still problems.  Perturbation 
theory does not seem to be converging for this quantity at low $x$ and $Q^2$. Indeed, there
have previously been suggestions that small-$x$ resummations may play an 
important role in $F_L(x,Q^2)$ \cite{lowxresum}. 
However, there also appears to be a problem at higher
$x$. The comparison of theory to data for $F_L(x,Q^2)$ is not satisfactory
for the higher-$x$ direct measurements of $F_L(x,Q^2)$.  Indeed, when we perform new global fits including
the $F_L$ data, we find $\chi^2$
of $44$ for the $36$ $F_L$ points at NLO, with a definite tendency for data to lie above 
theory. There is a big improvement at NNLO, with the corrections 
being large and positive, and the $\chi^2$ is $36/36$. However, there is still 
a tendency for data to lie a bit above theory. A comparison 
between data and theory is shown in Fig.\ref{fig:direct}.
In any case at high $x$ it 
is likely that higher twist is an important contribution to $F_L(x,Q^2)$.  

\section{The inclusion of Higher Twist}

The paucity of data for $F_L(x,Q^2)$, as compared with the case for $F_2(x,Q^2)$,
means that it is not possible to perform the entirely phenomenological analysis 
of higher twist that we performed for the latter \cite{MRSTtheory}, 
i.e. to include a higher
twist correction of the form $(c(x)/Q^2)F_2(x,Q^2)$ where we allowed $c(x_i)$ to be 
independent parameters representing 13 bins in $x$.\footnote{There is an
analysis for $F_L(x,Q^2)$ similar in spirit to this in 
\cite{Alekhin} which has only one $x<0.1$ bin. 
It also predates the calculation of the NNLO coefficient functions for 
$F_L(x,Q^2)$.}  
For $F_L(x,Q^2)$ the number of bins 
would have to be much smaller to avoid having gaps, or having only 1 or 2 points in a 
bin. Hence, we appeal to theoretical motivation for our 
choice of the higher-twist correction. For the case of the nonsinglet
higher-twist contribution to $F_2(x,Q^2)$, a correction 
of the form $(c(x)/Q^2)F_2(x,Q^2)$ has long been suspected to be enhanced by 
$1/(1-x)$ at high $x$. It also must satisfy the Adler sum rule, and hence vanish 
as $x \to 0$. The renormalon calculation \cite{renormalon} 
has exactly this trend, as indeed does the phenomenological higher twist
extracted by a global fit \cite{MRSTtheory}. In particular there seems to be no evidence for
a large contribution beyond the  nonsinglet contribution, and renormalon 
calculations  are problematical for such extra contributions \cite{Smye}.

For $F_L(x,Q^2)$ there is no reason to expect the same type of enhancement at 
high $x$, and also no reason for the higher-twist contribution to vanish at low $x$. In this case
a correction of the approximate form $(c/Q^2)F_2(x,Q^2)$ is expected, where
$c$ is constant. Again the renormalon calculation is in reasonable agreement 
with the naive prediction \cite{renormalonfl, renormalon}, giving a nonsinglet 
contribution of  
\begin{equation}
F^{HT}_L(x,Q^2) = \frac{A}{Q^2}(\delta(1-x)-2x^3)\otimes \sum_f Q_f^2 
q_f(x,Q^2).\label{eq:renormalon}
\end{equation}
This, and similar expressions are often presented with  
$F_2(x,Q^2)$ on the right-hand side. However, it is 
strictly the quarks that should appear, 
with the coefficient functions containing the 
appropriate quark-squared weighting, this combination of quark distributions 
and charge-squared weighting then being identical to the 
LO expression for $F_2(x,Q^2)$.  
The overall normalisation $A$ has been estimated to be $(8C_F\exp(5/3)/\beta_0) 
\Lambda_{QCD}^2$ \cite{renormalonfl}, but this is uncertain, and its value can also vary enormously by 
choosing $\Lambda_{QCD}$ defined at different orders. In \cite{Ringberg} the
generous NLO value\footnote{Note that the most recent MRST2004 values are 
$\Lambda_{QCD}^{\rm NLO}=347~\MeV$ and
$\Lambda_{QCD}^{\rm NNLO}=251~\MeV$.} of $\Lambda_{QCD}=347~\MeV$ was taken, giving a  
large higher-twist correction with $A=0.8~\GeV^2$.\footnote{In \cite{Ringberg}
$A$ was defined to be the coefficient of the whole higher-twist term for the 
first moment rather than the coefficient of the $x$-dependent function, hence 
the value of $A=0.4~\GeV^2$ was quoted.} 

\begin{figure}
\begin{center}
\centerline{\epsfxsize=0.66\textwidth\epsfbox{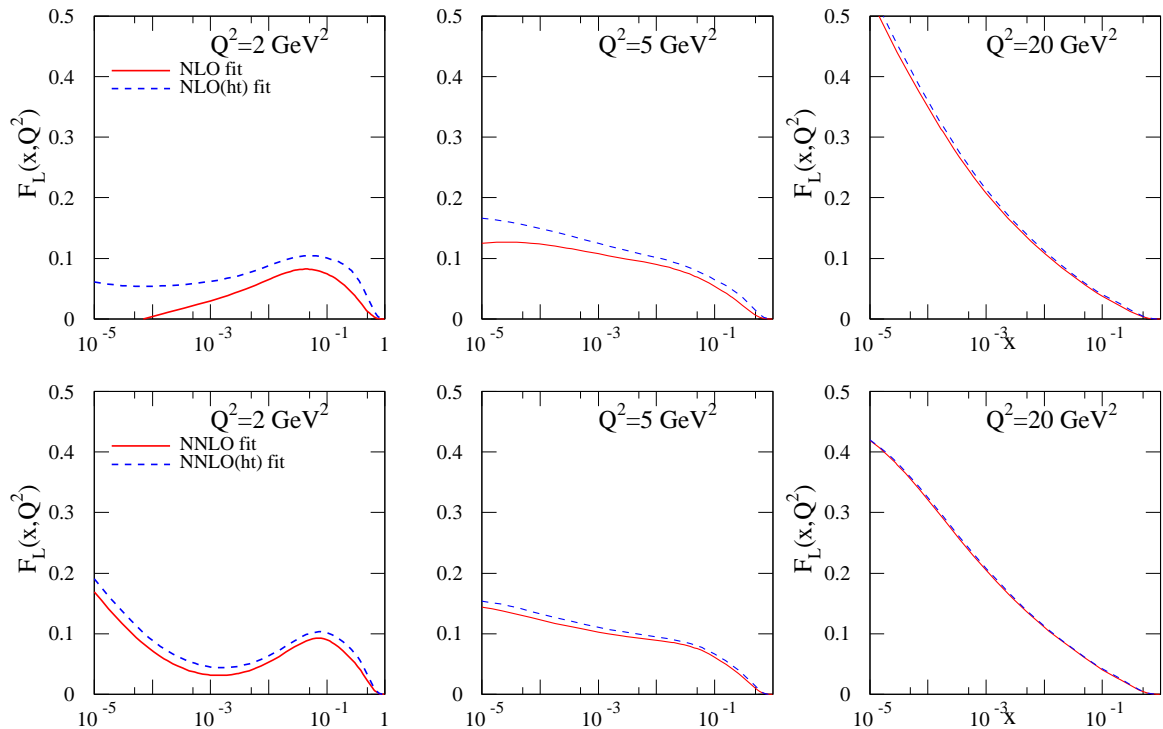}}
\caption{Predictions for $F_L(x,Q^2)$ at NLO (NNLO), with and without the renormalon correction, are shown in
the top (bottom) set of plots.}
\vspace{-1cm}
\label{fig:renorm}
\end{center}
\end{figure}

Here we perform global fits including both the direct data on $F_L(x,Q^2)$ and 
the indirect high $y$ data from HERA, in which we allow $A$ to be a free parameter. 
At both NLO and NNLO the quality of the fit to both the direct and the 
HERA data improves. For the direct data we get $\chi^2=34/36$ at NLO and 
one better at NNLO. The values $A=0.36$ and $A=0.16$ are found at NLO and NNLO respectively, each with an error of $\pm 0.08$. The quality of the fit to direct data is shown in 
Fig.\ref{fig:direct}, and the effect of the higher-twist contribution to $F_L(x,Q^2)$ can be
seen from Fig.\ref{fig:renorm}. The values 
of $A$ preferred by fitting only to the HERA data are larger ($A=0.58(0.25)$ at NLO(NNLO)), but this makes $F_L(x,Q^2)$ too large for the 
best fit to the direct data. 

We also tried a fit with a simpler form of the higher-twist 
correction
\begin{equation}
F^{HT}_L(x,Q^2) = \frac{A}{Q^2}\sum_f Q_f^2 
q_f(x,Q^2).\label{eq:renormalona}
\end{equation}
This lead to no changes of any significance. The fit is of essentially 
the same quality. The value of $A$ increases by a factor of 3, 
which is exactly what one gets if one takes the small-$x$ limit of 
Eq.(\ref{eq:renormalon}). This shows that except at very high $x$ 
the correction in Eq.(\ref{eq:renormalon}) reduces to that in 
Eq.(\ref{eq:renormalona}) with $1/3$ times the normalization. This is 
easily understood. In moment space the multiplicative factor in 
Eq.(\ref{eq:renormalon}) is $N/(N+2)$ (using the convention in 
\cite{renormalonfl}), which in the small-$x$ ($N\to 1$) limit becomes $1/3$.

We can also investigate the size of the renormalon correction by
further examination of the formulae in \cite{renormalonfl}. 
The renormalon correction is
obtained from taking the pole contributions of the 
inverse Borel transformation of Eq.(20) in 
\cite{renormalonfl}. This then gives the term in
Eq. (23) in \cite{renormalonfl}, but with an uncertain normalization. 
Alternatively one can derive
the perturbative expansion in the naive non-abelianization limit by
expanding Eq. (20) in powers of $s$, the Borel conjugate variable to $Q^2$,
and performing the inverse transformation term-by-term. 
Doing this in moment space for their $N=1$, i.e. the conventional $N=0$ 
(small-$x$ limit), we find, for $\alpha_S=0.36$, successive LO, 
NLO, NNLO,... contributions\footnote{To be precise, the series is
the perturbative expansion of the $N=1$ moment of the quark coefficient 
function. Thus, the moment of the quark contribution to $F_L$ (which should 
be dominant at highish $x$, and is in practice important at all $x$ at low 
$Q^2$) is the sum of the series multiplied by the moment of the quark
distribution.}
\beq
0.0765+0.052+0.043+0.043+0.050+...~~~,
\eeq
i.e. the LO coefficient function for $\alpha_S=0.36$ is $0.0765$, 
the NLO
correction in the large $\beta_0$ limit is $0.052$ the NNLO contribution in 
the same limit is $0.043$ {\it etc}.
Hence, for $\alpha_S=0.36$, which 
corresponds to $Q^2 \sim 2~\GeV^2$, the NNLO and NNNLO term are roughly the same and we
are led to keep three perturbative terms, with $0.043$ being an estimate of the higher
twist. Indeed, if one evaluates the finite part, i.e. the principal value, of Eq.(23) in \cite{renormalonfl} one gets $0.1782$ (to be compared with $0.0765+0.052+0.043=0.1715$) as the perturbative
contribution with the renormalon correction taken out, which is very
consistent.
         
Repeating this procedure for $\alpha_S=0.3$, i.e. $Q^2 \sim 4~\GeV^2$, one gets
\beq
0.064+0.036+0.025+0.021+0.020+0.023+...~~~.
\eeq         
Now ${\rm N}^3$LO is still smaller than NNLO, and perhaps ${\rm N}^4$LO is
representative of the overall higher-twist contribution, i.e. $0.02$, which is about half of
$0.043$, consistent with a $1/Q^2$ dependence when $Q^2$ goes from $2 \to 4~\GeV^2$. In this case, explicit
evaluation of the principle value of Eq.(23) gives $0.1482$, to be compared with 0.146 
(the sum of the first four terms). Again we have consistency, but where the series should be truncated is a
function of $Q^2$. However, the term one includes at higher $Q^2$ is rather
small.

Since the renormalon term is $A/Q^2\times N/(N+2)$ then at $N=1$ 
our higher twist should be
$A/(3Q^2)$.  For $Q^2=2~\GeV^2$ we therefore have $A/6 \sim 0.04$, that is $A \sim 0.24$.
This should be compared with the values $A=0.38$ and $A=0.16$ obtained in our
NLO and NNLO fits. However, it is clear that the NLO value is too high because it
is missing significant NNLO corrections, whereas perhaps all the
correction to NNLO should be higher twist. In this case the fitted value of $0.16$ compares well
with the approximate prediction of $0.24$, especially
considering that the fit to HERA data alone favours $A \sim 0.25$.

Considering the variation with $N$, or equivalently with $x$, leads to 
complications however. The higher-twist correction at $N=1$ (that is the conventional $N=0$) is
$A/(3Q^2)$, whereas at larger $N$ it tends to $A/Q^2$. This might suggest that
at a fixed $Q^2$ a particular term in the series, e.g NNLO at 
$\alpha_S=0.36$,
increases with increasing $N$, saturating at 3 times its $N=1$ value at large
$N$. This is not the case. In fact the NNLO term is slowly varying with $N$,
and actually decreases for very large $N$. For example, for $N=6$ and
$\alpha_S=0.36$ we get
\beq
0.0218+0.0297+0.0404+0.056+.....
\eeq
i.e. the NNLO term is slightly smaller than at $N=1$; but note that we now have no
convergence. We have to go to higher $Q^2$ to get any evidence of
convergence. If we take, for example, $\alpha_S=0.24$ (corresponding to $Q^2 \sim 10~\GeV^2$) then 
for $N=6$ we have
\beq
0.0145+0.0132+0.0120+0.0111+0.0109+0.0114+...~~~.
\eeq
So at this scale there is some convergence. Compare this to the same
$Q^2$ for $N=1$,
\beq
0.0508+0.0233+0.0129+0.0085+0.0066+0.0061+0.0065+...~~~.
\eeq   
Here we would keep 5 or 6 terms in the series and the higher-twist contribution is
0.0065 (compared to $0.24/(3Q^2)=0.008$, which is not too bad). However at $N=6$ we would keep 4
or 5 terms and the higher twist is $0.011$. So at the higher $Q^2$ the picture
is reasonably consistent, though arguably we should treat more terms as
perturbative at low $x$ than at high $x$, and the higher-twist term enters at about 6th
or 7th order. On the other hand, at low $Q^2$ we cannot say that the higher twist is
roughly NNNLO. This is the case at low $x$, but at high $x$ the situation is confused. If we evaluate the inverse transformation explicitly for $N=6$ and
$\alpha_S=0.36$ we find $0.022$ for the perturbative contribution, 
which implies keeping something like just the
LO term with NLO representing the size of the higher twist, two orders
lower than at lower $N$ and $x$. However the size of the higher twist is then not
quantitatively consistent. We conclude
that the high $x$ and low $Q^2$ domain is ``dangerous''. This is another reason, along with
target mass, to avoid fitting data in this kinematic region. For $\alpha_S=0.24$ (that is $Q^2 \sim 10~\GeV^2$) the explicit integral
gives $0.058$ for $N=6$, consistent with keeping the first five terms.
For $N=1$ it gives $0.0987$, again roughly corresponding to the first 5 terms. Hence, as long as
we stay away from high $N$ combined with low $Q^2$ everything is consistent.
Nevertheless, at high $N$ and low $Q^2$ one can still get some idea of what is happening, i.e.
the whole perturbative expansion is difficult to distinguish from higher
twist, but the problem is that there is no quantitative approach applicable in
this (high $x$, low $Q^2$) domain.

\section{Conclusions}

\begin{figure}
\begin{center}
\centerline{\epsfxsize=0.7\textwidth\epsfbox{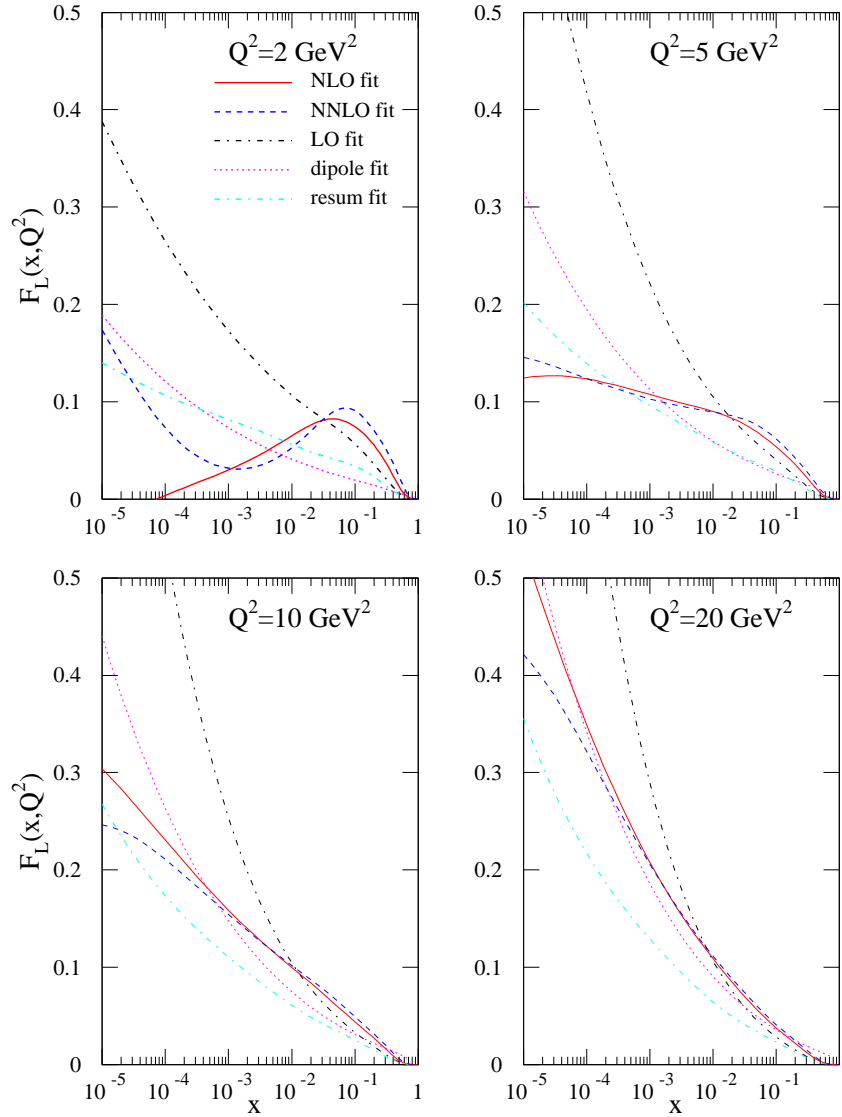}}
\vspace{-0.3cm}
\caption{$F_L(x,Q^2)$ predicted from the global fit at LO, 
NLO and NNLO, also from a fit which performs a double resummation 
of leading $\ln(1/x)$ and $\beta_0$ terms, and finally from a dipole 
model type fit.}
\vspace{-1cm}
\label{fig:theory}
\end{center}
\end{figure}

There is much to learn from the inclusion of the data which are both directly, and
indirectly, sensitive to $F_L(x,Q^2)$ in a global fit. There is clear 
evidence that NLO in perturbation theory alone is not a sufficiently 
sophisticated approach to fit either the direct or indirect data. There is an 
improvement from including the recently calculated NNLO corrections, but 
still some significant problems. The further inclusion of a higher-twist 
contribution, inspired by the renormalon correction in the nonsinglet sector,
is very successful, and the normalization of this contribution seems to be 
consistent with the theoretical expectations. However, there is still
room for further theoretical corrections, particularly at low $x$ and $Q^2$. 
There are numerous suggestions that even higher orders in perturbation theory are important here 
due to the large $\ln(1/x)$ terms. Indeed, the NNNLO coefficient functions
calculated in $\msb$ scheme in the latter paper of \cite{NNLOfl} lead to large 
corrections for $Q^2=2~\GeV^2$ and $x\sim 0.0001$, and studies 
including resummations suggest important modifications in this region 
for $F_L(x,Q^2)$, see, for example, \cite{resum, resumcdw}. 
The same is true for studies using dipole models \cite{gbw, RSTdipole} which
contain both higher-order corrections and higher twists (different from those 
in the renormalon correction considered here). A variety of predictions from 
different theoretical approaches is shown in Fig.\ref{fig:theory}. A direct 
measurement of $F_L(x,Q^2)$ for $Q^2 \lapproxeq 5~\GeV^2$ would be very important in determining which 
theoretical approaches are most reliable, as discussed in detail in 
\cite{Ringberg}.

\section*{Acknowledgements}

RST thanks
the Royal Society for the award of a University Research Fellowship and ADM 
 thanks the Leverhulme Trust for the award of an Emeritus
Fellowship. The IPPP gratefully acknowledges financial support from the UK
Particle Physics and Astronomy Research Council. We would also like to thank 
Dick Roberts for many years of fruitful and enjoyable collaboration on the 
topic of parton distributions and for setting in place much of the framework
which enables us to perform the global analyses.\\

\vspace{-0.6cm}


\end{document}